\documentclass[conference]{IEEEtran}

\usepackage{cite}
\usepackage{amsmath,amssymb,amsfonts}
\usepackage{algorithm}
\usepackage{algorithmic}
\usepackage{hyperref}
\usepackage{graphicx}
\usepackage{textcomp}
\usepackage{xcolor}
\def\BibTeX{{\rm B\kern-.05em{\sc i\kern-.025em b}\kern-.08em
    T\kern-.1667em\lower.7ex\hbox{E}\kern-.125emX}}
\begin{document}

\title{Join Cardinality Estimation with OmniSketches}

\author{\IEEEauthorblockN{David Justen}
\IEEEauthorblockA{TU Berlin \& BIFOLD \\
david.justen@tu-berlin.de}
\and
\IEEEauthorblockN{Matthias Boehm}
\IEEEauthorblockA{TU Berlin \& BIFOLD \\
matthias.boehm@tu-berlin.de}
}

\maketitle
\thispagestyle{plain}
\pagestyle{plain}

\begin{abstract}
Join ordering is a key factor in query performance, yet traditional cost-based optimizers often produce sub-optimal plans due to inaccurate cardinality estimates in multi-predicate, multi-join queries. Existing alternatives such as learning-based optimizers and adaptive query processing improve accuracy but can suffer from high training costs, poor generalization, or integration challenges.
We present an extension of OmniSketch---a probabilistic data structure combining count-min sketches and K-minwise hashing---to enable multi-join cardinality estimation without assuming uniformity and independence. Our approach introduces the OmniSketch join estimator, ensures sketch interoperability across tables, and provides an algorithm to process alpha-acyclic join graphs. Our experiments on SSB-skew and JOB-light show that OmniSketch-enhanced cost-based optimization can improve estimation accuracy and plan quality compared to DuckDB. For SSB-skew, we show intermediate result decreases up to 1,077x and execution time decreases up to 3.19x. For JOB-light, OmniSketch join cardinality estimation shows occasional individual improvements but largely suffers from a loss of witnesses due to unfavorable join graph shapes and large numbers of unique values in foreign key columns.
\end{abstract}

\section{Introduction}

Query optimization is a critical problem in database systems that has been investigated for many decades~\cite{DBLP:conf/sigmod/BabcockC05, DBLP:conf/vldb/OnoL90, DBLP:conf/sigmod/SelingerACLP79}. For relational analytics workloads, join ordering contributes heavily to the overall performance as query plans with sub-optimal join orders can be penalized with execution time increases of multiple orders of magnitude~\cite{DBLP:journals/vldb/LeisRGMBKN18}.

\textbf{Traditional Query Optimization:} Database systems traditionally employ cost-based query optimization~\cite{DBLP:conf/sigmod/SelingerACLP79}. In this process, the optimizer enumerates sub-plans of the query, assigns a cost to each sub-plan by estimating its cardinality, and finally combines them to find the query plan with the lowest cost. One of the key goals of join order enumeration is to minimize the number of intermediate results. During cardinality estimation, optimizers often rely on database statistics and sketches such as histograms. However, for multi-predicate and multi-join queries, these techniques resort to assuming data uniformity and independence, leading to consistent under-estimations and sub-optimal plan generation~\cite{DBLP:journals/pvldb/LeisGMBK015}.

\textbf{Learning-based Optimization:} To address these limitations, two major research directions have emerged. Learning-based optimizers~\cite{DBLP:journals/pvldb/MarcusNMZAKPT19, DBLP:conf/sigmod/YangC0MLS22, DBLP:conf/cidr/KipfKRLBK19, DBLP:journals/pvldb/YangLKWDCAHKS19} have been introduced to improve query plan quality. These approaches train machine learning models on the database and sample queries and use the model to predict query plans for incoming queries. While learning-based optimizers have shown improvements over traditional optimizers, some of them require extensive initial training phases, and do not generalize well for unknown queries~\cite{DBLP:journals/pvldb/ZhangCPR23}.

\textbf{Adaptive Query Processing:} As an alternative line of research, numerous adaptive query processing (AQP) techniques have been introduced over the last two decades~\cite{DBLP:conf/sigmod/HellersteinA00, DBLP:conf/cidr/BabuB05, DBLP:journals/ftdb/DeshpandeIR07, DBLP:journals/pvldb/JustenRFLTLBHZMB24}. Instead of relying on cardinality estimates during query optimization, AQP approaches gather statistics such as selectivities during query execution and continuously adapt the query plan based on the collected telemetry. Although many of these techniques have shown promising results, adaptive join re-ordering approaches are often difficult to integrate into existing systems or make substantial applicability concessions~\cite{DBLP:journals/pvldb/JustenRFLTLBHZMB24}.

\textbf{OmniSketch Cardinality Estimation:} While learning-based optimization and AQP are active areas of research, new sketches that may improve traditional query optimization are also emerging. The OmniSketch~\cite{DBLP:journals/pvldb/PunterPG23} combines count-min sketches~\cite{DBLP:journals/jal/CormodeM05} with K-minwise hashing~\cite{DBLP:conf/pods/PaghSW14} and allows for multi-attribute cardinality estimates with probabilistic error guarantees. In this work, we extend the OmniSketch to enable cardinality estimation for multi-join queries and assess the extended OmniSketch in the context of cost-based query optimization to examine its trade-offs.

\textbf{Contributions:} Our primary contribution is the extension of OmniSketches for multi-join cardinality estimation. The code is open-source as a C++ library on GitHub\footnote{C++ library available at \url{https://github.com/d-justen/OmniSketchCpp}.}. Our detailed contributions are the following:

\begin{itemize}
    \item We introduce an OmniSketch design adaption that ensures the multi-table interoperability via primary and foreign key constraints (Section~\ref{sketch-interoperability}).
    \item We define a strategy for join cardinality estimation that reformulates joins to sampled set membership predicates and a second approach that makes use of secondary sketches (Section~\ref{join-concept}).
    \item We contribute an algorithm to derive OmniSketch operation plans from alpha-acyclic join graphs for multi-join cardinality estimates (Section~\ref{multi-join}).
    \item With the JOB-light and SSB-skew benchmarks, we systematically study the benefits, shortcomings, and trade-offs of cost-based query optimization with OmniSketches as opposed to DuckDB (Section~\ref{experiments}).
\end{itemize}

\section{Background}
\label{background}

\begin{figure}[!t]
    \centering
    \includegraphics[width=0.85\linewidth]{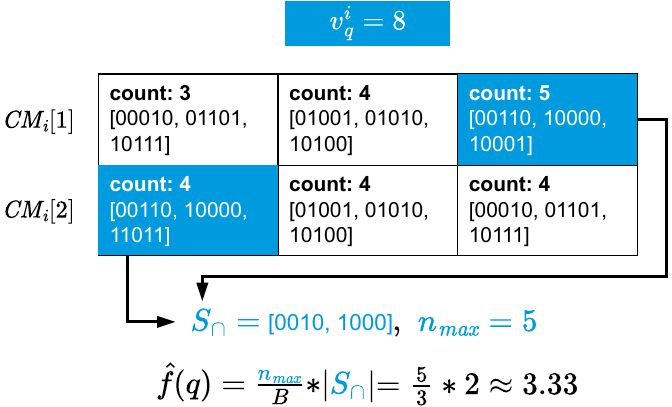}
    \vspace{-0.2cm}
    \caption{OmniSketch Overview. An Omnisketch $\mathit{CM_i}$ with depth $d = 2$, width $w = 3$, and min-hash sample size $B = 3$. To estimate the count-aggregate for a predicate value $v^i_q = 8$, we determine the cell index $k$ for each row $\mathit{CM_i}[j], 1 \leq j \leq d$ by hashing $k =h_j(v^i_q) \mod w$, and compute $S_{\cap}$ by intersecting the contained record id hashes. The estimate $\hat{f}(q)$ is determined by scaling the intersection cardinality with the maximum record count $n_{\text{max}}$ divided by the min-hash sample size $B$.}
    \label{fig:omnisketch-overview}
    \vspace{-0.2cm}
\end{figure}

The OmniSketch~\cite{DBLP:journals/pvldb/PunterPG23} addresses a critical limitation in streaming synopses: traditional sketches such as Bloom filters~\cite{DBLP:journals/cacm/Bloom70}, count-min sketches~\cite{DBLP:journals/jal/CormodeM05}, or HyperLogLog~\cite{hyperloglog} are designed for single-attribute aggregates and do not support multi-attribute predicates. OmniSketch is a novel sketch designed to provide count-aggregates over complex, high-velocity streams with point and range predicates on arbitrary attributes. The sketch provides probabilistic error bounds and a tunable space–accuracy trade-off. Figure~\ref{fig:omnisketch-overview} gives an overview of the sketch structure and the cardinality estimation process.

\textbf{Sketch Structure:} OmniSketches internally consist of a count-min sketch with $d$ rows, $w$ cells per row and one hash function $h_j$ per row. An OmniSketch $\mathit{CM_i}$ is built for each searchable attribute $a_i \in A$ of a single table. The sketches have a counter $\mathit{cnt}_i[j,k]$ in each cell $\mathit{CM_i}[j,k]$ (analogous to count-min sketches) but also store a sample of record ids $S_i[j,k]$ using K-minwise hashing\cite{DBLP:conf/pods/PaghSW14} with a sample size of $B$. To insert a record $R$ with attribute values $r_i$ and a record id $r_0$, for each $j \in \{1, \cdots,d\}$ we determine a cell $\mathit{CM_i}[j, k]$ by hashing the attribute with $k = h_j(r_i) \mod w$. We increment its record count $\mathit{cnt}_i[j, k]$, compute a record id hash $g(r_0)$ and add it to the sample $S_i[j, k]$ if $\mathit{cnt}_i[j, k] < B$. Otherwise, if $g(r_0) < \mathit{max}(S_i[j, k])$, we replace $\mathit{max}(S_i[j, k)])$ with $g(r_0)$.

\textbf{Cardinality Estimation:} To estimate the cardinality under a given predicate value $v^i_q$, we hash the value to find $d$ OmniSketch cells, their record counts and record id samples. We set $n_{\mathit{max}}$ to the maximum record count in these matches and compute the sample intersection $S_\cap = \bigcap\limits_{1 \leq j \leq d} S_i[j, h_j(v^i_q)]$. By scaling up the size of the intersection, we compute the cardinality estimate: $\hat{f}(q) = n_{\mathit{max}} /B * |S_\cap|$. Multi-attribute cardinality estimation follows the same logic. We probe each $v^i_q \in V_q$ into $\mathit{CM_i}$, compute $n_{\mathit{max}}$ from all matching cells, and compute the intersection of $|V_q| * d$ samples.

\section{Join Extension}

In order to extend the OmniSketch structure for multi-table cardinality estimation, we create OmniSketches for all searchable attributes in each table and assume the availability of a primary key column in each of them, so that the min-hash samples are built on each table's primary keys. This design allows us to retrieve primary key hashes from single-table queries. As it is not possibly by default to probe these hashes into an OmniSketch on a foreign key column, we unify the OmniSketch hashing strategy to allow for interoperability. With the extended OmniSketch, we devise and discuss a strategy to estimate one-to-many joins by probing primary key hashes. We also introduce an alternative strategy for better accuracy and lower latency with secondary OmniSketches. Finally, we define OmniSketch inter-table operations and introduce a join graph traversal algorithm that produces operation sequences to estimate multi-join queries.

\subsection{Sketch Interoperability}
\label{sketch-interoperability}

In order to make OmniSketches interoperable so that an OmniSketch on a foreign key column can be probed with a primary key hash, we unify the hash functions used in all OmniSketches. To that end, we base all hash functions on a single 64-bit hash function $g^\prime(r_i)$. The min-hash samples contained in the OmniSketch cells use only that hash function. To determine the OmniSketch cells in $\mathit{CM_i}$ for a given value $r_i$, we compute $g^\prime(r_i)$ and split the result into two 32-bit hashes $g^\prime_1, g^\prime_2$~\cite{modernbloom}. With these partial hashes, we construct a hash function for each row in $\mathit{CM_i}$ as $h_j(r_i) = (g^\prime_1 + j g^\prime_2)$~\cite{DBLP:conf/esa/KirschM06}. This adaption results in two different probing methods. For attribute values we compute the hash $g^\prime(v^i_q)$ and split it to find the cell indices in each row, and for a hashed primary key we skip the hashing and split the hash directly.

\subsection{Join Cardinality Estimation}
\label{join-concept}

We introduce two strategies for single join estimation: a universal strategy that estimates joins as sampled set membership predicates and an alternative strategy that builds secondary sketches for additional accuracy and lower latency.

\textbf{PK Sample Joins:} For the PK sample join strategy, we treat joins like set membership predicates (e.g., \texttt{a IN (1, 2, 3}), in which each join key $v^i_q \in V^i_q$ is part of the set to be probed. We can estimate the cardinality of such a query by probing each key individually into an OmniSketch and summating the estimates. In the OmniSketch join estimation case, we do not know the full set of primary keys. Consider a join query such as \texttt{SELECT count(*) FROM R, T WHERE R.sid = T.id And T.a = 3} with a primary key constraint on \texttt{T.id}. We probe the OmniSketch on \texttt{T.a}, producing a cardinality estimate for $\mathtt{'T.a = 3'}$ and a min-hash sample $S_\cap$ on \texttt{T.id}. We compute the sampling probability $p = |S_\cap| / \hat{f}(\mathtt{'T.a = 3'})$ and probe the samples into the OmniSketch on \texttt{R.sid}. Given an $n^i_\mathit{max}$ and $S^i_\cap$ for each sample probe, we estimate the cardinality and scale it up with:

\begin{equation*}
    \hat{f}(q) = \frac{1}{p}\sum\limits_{1 \leq i \leq |S_\cap|} \frac{n_{\mathit{max}}^i}{B} |S^i_\cap|
\end{equation*}

For multi-join support, we compute the union of all samples $S_\cup^i$, and store them with their respective $n^i_{\mathit{max}}$ in a map. That map is used as an intermediate result that can be intersected with other min-hash samples on the same primary key. By associating each sample with its $n^i_{\mathit{max}}$, we can compute a single n-way intersection with other min-hash sample unions instead of intersecting each probe result individually with other predicate or join estimation results.
This strategy introduces the ability estimates to estimate join cardinalities but also comes with substantial shortcomings. 
Since the primary key hashes of \texttt{T.id} are uncorrelated with the minimal hashes of \texttt{R.id}, they act as a random sample. Thus, the upscaling with the sampling probability introduces an assumption that the sample is representative for all qualifying primary keys. However, this may not be the case if the data distribution in the foreign key column is non-uniform. In the worst case, the foreign key column could contain a heavy hitter that is also a qualifying primary key, leading to severe under-estimation. Another drawback of this method is its high compute cost as each individual probe requires a multi-way intersection. Finally, for join graphs in which a fact table joins multiple dimension tables, the estimation relies on the intersection of multiple random primary key samples. As the probability for a match to be in all random samples can become very low (i.e., the product of individual sampling probabilities), these join graphs are likely to run out of witnesses. 

\textbf{Secondary Sketches:} For faster join estimation with higher accuracy, we can build \textit{secondary} OmniSketches on dimension tables that map attribute values of the dimension tables directly to their corresponding primary keys of a fact table. For that, we first build the OmniSketches for the fact table. Once these sketches are built, we create the dimension sketches. Instead of inserting their primary keys into the min-hash samples, we probe each primary key into the referencing foreign key column sketches, insert the resulting $S_\cap$ sample into the dimension side samples and increment their record counts by the cardinality estimate. Note that the resulting secondary OmniSketch suffers some information loss (as opposed to a sketch that could be built by probing a hash table on the foreign key column). However, our experiments show that this effect is negligible for cardinality estimation accuracy, and it can be further reduced by increasing the sample size for the primary sketch. Building secondary sketches trades a higher upfront sketch buildup time with a lower estimation latency as all joins and predicates on a fact table and its dimension tables only require a single multi-way intersection. Moreover, this approach resolves the previous method's problem of assuming representativeness as all OmniSketches contain min-hash samples on the same primary key column.

\begin{figure}
    \centering
    \includegraphics[width=0.5\linewidth]{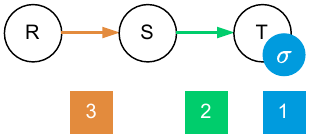}
    \vspace{-0.2cm}
    \caption{Probe Sequence Example. Given a join graph with directed edges of foreign key columns pointing to primary key columns and a predicate on \texttt{T}. The only legal sequence of OmniSketch probes is (1) the predicate on \texttt{T}, (2) probing the resulting primary key hashes into \texttt{S}, and (3) probing the resulting \texttt{S.id} hashes into \texttt{R}.}
    \vspace{-0.2cm}
    \label{fig:sequence-example}
\end{figure} 

\subsection{Join Graph Traversal}
\label{multi-join}
While multi-join cardinality estimation with secondary sketches works analogous to single-table table estimation, using these sketches is not always possible. If the database is not organized in a star-schema, we resort to the default join sampling technique. Estimating arbitrary alpha-acyclic join graphs requires computing a legal sequence of OmniSketch probe operations due to the directional constraint that probing \texttt{T.id} primary keys into a \texttt{R.sid} OmniSketch yields a min-hash sample of \texttt{R.id}. Therefore, all additional filters on \texttt{T} must be applied to \texttt{T.id} before probing into \texttt{R.sid}. An illustrative example for this problem can be found in Figure~\ref{fig:sequence-example}.

\begin{figure}
    \centering
    \includegraphics[width=0.482\linewidth]{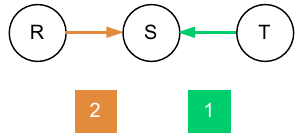}
    \vspace{-0.2cm}
    \caption{Expansion Example. Given a join graph, where foreign keys of \texttt{R} and foreign keys of \texttt{T} join with primary keys of \texttt{S}. In that case, neither sequence (1), (2) or (2), (1) is legal for cardinality estimation with regular OmniSketch probes. With a primary key expansion, we can estimate any join first, and then perform a join probe on the remaining relation.}
    \vspace{-0.2cm}
    \label{fig:expansion-example}
\end{figure} 

\textbf{Primary Key Expansions:} In certain situations, we must loosen the sequence constraint. Consider the directed join graph from Figure~\ref{fig:expansion-example}, in which foreign key sides point to the primary key side. As \texttt{S.id} must be probed into \texttt{R.sid} as well as \texttt{T.sid}, we either produce a sample on \texttt{R.id} that cannot be used to probe \texttt{T.sid} or a sample on \texttt{T.id} that cannot probe the OmniSketch on \texttt{R.sid}. To resolve this stalemate, we introduce a second kind of OmniSketch join estimator: the primary key expansion. For this operation, we perform the PK sample join and intersect the result with any other predicate or join probe results on the foreign key side. However, instead of returning the resulting hashes, we filter the input sample based on whether it has led to matches or not. In our example, we would expand the \texttt{S.id} hashes on \texttt{R.sid}, resulting in a filtered min-hash sample on \texttt{S.id} with a cardinality estimate for $\mathtt{R} \bowtie \mathtt{S}$. After that, we include \texttt{T} in the estimate with a PK sample join of the filtered sample into \texttt{T.sid}. Note that also a secondary sketch, if available, can be used either for the expand step or the probe step.

\begin{algorithm}
\caption{Join Graph Traversal Algorithm Overview. The algorithm iteratively merges join graph ears and expands primary keys whenever necessary.}
\label{fig:traversal-algo}
\begin{algorithmic}[1]
\renewcommand{\algorithmicrequire}{\textbf{Input:}}
\renewcommand{\algorithmicensure}{\textbf{Output:}}
\REQUIRE Join Graph $G \gets (V, E)$
\ENSURE Cardinality Estimate
\WHILE{$|\mathit{G.V}| > 1$}
    \FORALL{$v_i \in \mathit{G.V}$}
        \IF{$\{\exists e \in \mathit{G.E} | v_i = \mathit{e.fk}\}$}
        \STATE \textbf{continue}
        \ENDIF
        \STATE $E^{v_i} \gets \{ e \in \mathit{G.E}|v_i = \mathit{e.pk}\}$
        \IF{$|E^{v_i}| = 1$}
            \STATE $S_\cap^{v_i} \gets \mathit{EstimatePredicates}(v_i)$
            \STATE $\mathit{AddSetPredicate}(e_1^{v_i}.\mathit{fk}, S_\cap^{v_i})$
            \STATE $\mathit{G.V} \gets \mathit{G.V} \setminus v_i$
        \ELSE
            \STATE $E^{v_i} \gets \{ e \in E_{v_i} | \mathit{e.fk} \textit{ has exactly one edge}\}$
            \IF{$E^{v_i} = \emptyset$}
                \STATE \textbf{continue}
            \ENDIF
            \STATE $S_\cap^{v_i} \gets \mathit{EstimatePredicates}(v_i)$
            \STATE $\mathit{AddExpansion}(v_i, \mathit{Expand}(e_1^{v_i}.\mathit{fk}, S_\cap^{v_i}))$
            \STATE $\mathit{G.V} \gets \mathit{G.V} \setminus e_1^{v_i}.\mathit{fk}$
        \ENDIF
        \STATE $\mathit{G.E} \gets \mathit{G.E} \setminus e_1^{v_i}$
    \ENDFOR
\ENDWHILE
\RETURN $\mathit{EstimatePredicates}(v_1)$
\end{algorithmic}
\end{algorithm}

\textbf{Traversal Algorithm:} Algorithm~\ref{fig:traversal-algo} gives an overview of the graph traversal strategy for multi-join cardinality estimation, inspired by the GYO ear removal algorithm~\cite{DBLP:conf/compsac/YuO79}. Instead of hypergraphs, it operates on a directed join graph $G \gets (V, E)$ with relations $V$ and joins $E$, connecting a primary key side $\mathit{e.pk}$ and a foreign key side $\mathit{e.fk}$. We also expect to know the predicates on each $v \in \mathit{G.V}$ and evaluate them with an $\mathit{EstimatePredicate}$ method. The strategy of the algorithm is to find graph "ears" that are only connected to one other relation via their primary key column (line 3-7). We remove those ears by applying them as set membership predicates to the foreign key side (line 9). If a relation has multiple joins on its primary key column (line 12), we use the primary key expansion on any foreign key side that does not have other edges (line 13-17) and remove the foreign key side (line 18). The depicted algorithm only provides an overview and does not include processing steps for join graphs containing rings. However, these rings can be processed as well by applying a join predicate to the foreign key side and remove the common edge, if the node is reachable through other edges of the primary key side.

\textbf{Running Out of Witnesses:} If we run out of witnesses during the join graph traversal, we fallback to heuristics to compute a cardinality estimate. For simple predicates, set membership predicates, and sample joins, we set the cardinality estimate for each probe with an empty result to $n_{\mathit{max}} / B$, which is the minimal cardinality estimate the OmniSketch would have been able to give for a single matching hash. If we run out of witnesses while intersecting multiple join or predicate results, we multiply the associated selectivities. With these heuristics, we pick up the common assumptions of data uniformity and independence but are less likely to do so for plans with large cardinalities as these have a lower probability of running out of witnesses.

\section{Experiments}
\label{experiments}
Our experimental evaluation studies the performance and accuracy of join cardinality estimation with OmniSketches. To this end, we integrate our join graph traversal algorithm with DPsize and enumerate join plans for the SSB-skew~\cite{DBLP:journals/pvldb/JustenRFLTLBHZMB24} benchmark on scale factor 100 and the JOB-light benchmark~\cite{DBLP:conf/cidr/KipfKRLBK19}. We compare the resulting plans with the default plans from DuckDB~\cite{DBLP:conf/sigmod/RaasveldtM19} v1.2.2. For the SSB-skew dataset, we set the min-hash sample size $B \gets 128$, depth $d \gets 3$, fact table width $w_f \gets 256$, and dimension table width $w_d \gets 32$. We build OmniSketches on each attribute used in predicates and joins, resulting in a total estimated space consumption of about 4.15\,MiB for all primary sketches and 4.93\,MiB with secondary sketches included. For the JOB-light, we use depth $d \gets 3$, width $w \gets 256$ and $B \gets 256$ across all tables on all attributes used in the benchmark, resulting in a total estimated size of 12.1\,MiB. We do not build any secondary sketches for JOB-light, as all joins in the benchmark are performed on the primary key of the \texttt{title} table. All benchmarks are executed on a Macbook Pro M3 Max with 36\,GiB of main memory.

\begin{figure}
    \centering
    \includegraphics[width=\linewidth]{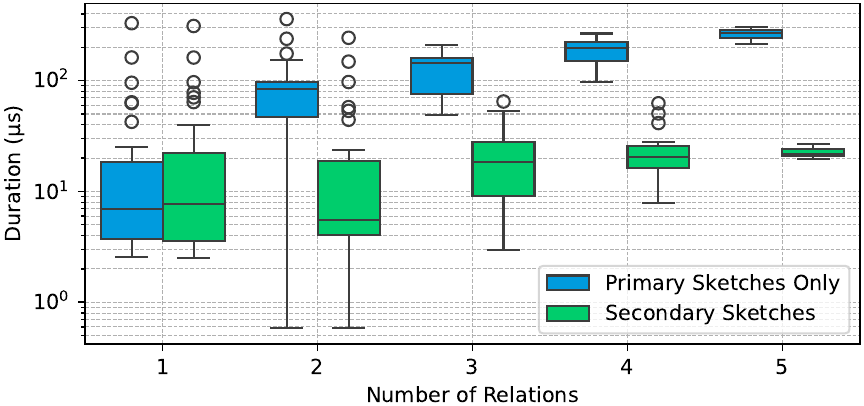}
    \vspace{-0.5cm}
    \caption{Cardinality Estimation Latency by Number of Relations in Sub-plan on SSB-skew.}
    \vspace{-0.2cm}
    \label{fig:duration-boxplot}
\end{figure}

\textbf{Estimation Latency:} Figure~\ref{fig:duration-boxplot} shows the latencies for query sub-plan cardinality estimation on SSB-skew, grouped by the number of relations per sub-plan. We compare the durations for the case, in which we only use primary OmniSketches with the PK sample join strategy with the case with secondary sketches enabled. While the primary-only case shows substantially longer estimation times for 2+ relations, the secondary sketch case only mildly deteriorates in performance for each additional join. The longest measured end-to-end estimation time for a whole query was 1.87\,ms in the primary-only case and 0.59\,ms with secondary sketches. We omit the individual sub-plan estimation times for JOB-light as they are similar to the primary-only case of SSB-skew. However, the longest end-to-end estimation time for JOB-light was 7.02\,ms. We conclude that estimation times are neglibible if secondary sketches can be used. For queries with larger numbers of joins, the estimation overhead may become notable if query processing is cheap and only primary sketches are available.

\begin{figure}
    \centering
    \includegraphics[width=\linewidth]{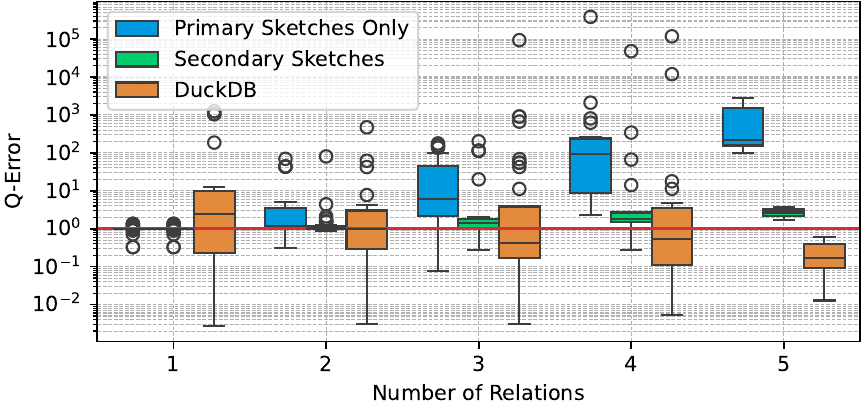}
    \vspace{-0.5cm}
    \caption{Q-Error by Number of Relations in Sub-plan on SSB-skew.}

    \label{fig:ssbskew-qerror-boxplot}
\end{figure} 

\begin{figure}
    \centering
    \includegraphics[width=\linewidth]{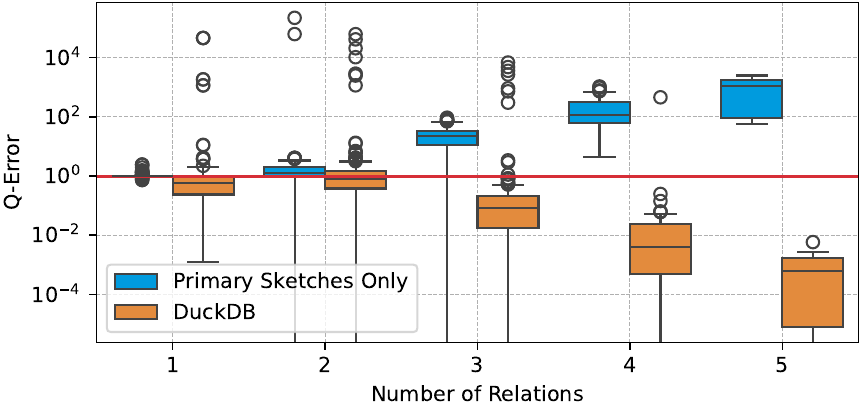}
    \vspace{-0.5cm}
    \caption{Q-Error by Number of Relations in Sub-plan on JOB-light.}
    \vspace{-0.2cm}
    \label{fig:joblight-qerror-boxplot}
\end{figure} 

\textbf{Estimation Error:} We measure the Q-Error to determine the quality of the OmniSketch cardinality estimation and compare it with DuckDB's default cardinality estimator. We define the Q-Error as $\mathit{CardEst} / \mathit{ActualCard}$ to differentiate between under-estimation (Q-Error $<$ 1) and over-estimation (Q-Error $>$ 1). Figure~\ref{fig:ssbskew-qerror-boxplot} depicts the Q-Errors for all sub-plans enumerated in the SSB-skew benchmark, grouped by the number of relations in the sub-plan. The experiment using only primary OmniSketches vastly over-estimates sub-plans with more than three relations, partly because of hash collisions but also because the fallback heuristics tend to over-estimate sub-plans with negatively correlated predicates. With secondary sketches enabled, the amount of over-estimation can be drastically reduced. The DuckDB cardinality estimator suffers from over- and under-estimation but tends towards under-estimation for three or more relations.
Figure~\ref{fig:joblight-qerror-boxplot} shows the Q-Errors for sub-plans from the JOB-light benchmark. Note that we only compare primary OmniSketches in this experiment as all relations in the JOB-light benchmark join on the primary key of the \texttt{title} table. Effectively, these join graphs are processed with $n-1$ primary key expansions and one PK sample join for each sub-plan with $n$ relations. In this benchmark, the OmniSketch join estimation strategy systematically over-estimates sub-plans with three or more relations. As our approach frequently runs out of witnesses during primary key expand operations for these sub-plans, it resorts to the heuristics, which tend to over-estimate individual joins and propagate these over-estimations with an increasing number of relations. Nevertheless, the majority of absolute Q-Errors, especially for sub-plans with four or more relations is smaller than DuckDB's absolute Q-Errors.

\textbf{Plan Quality:} Finally, we examine the quality of query plans emitted from DPSize join enumeration with OmniSketches. For that, we run SSB-skew (using secondary sketches) and JOB-light on DuckDB with eight threads. We execute the plans generated from DuckDB's optimizer and our plans in DuckDB and measure end-to-end execution times and the cumulated join cardinalities ($C_{\mathit{out}}$). For the OmniSketch plan execution times, we also add the individual cardinality estimation latencies. Table~\ref{tbl:plan-quality} shows a summary of the experiment results. For SSB-skew, the total execution time decreases from 9.16 seconds to 6.02 seconds, with a maximum execution time improvement of 3.19x. The OmniSketch approach reduces the total number of intermediates from 2.6 billion to 367 million with a maximum decrease of 1,077x. However, for JOB-light, the experiment shows an increase in total execution time from 10.65 seconds to 12.46 seconds, even though the number of total intermediates is only slightly larger and most queries show a slight decrease of intermediate results. One of the reasons for this deterioration is a single query that has a much larger execution time than all other queries of the benchmark. This query produces 4\,\% more intermediates with OmniSketches but deteriorates in performance from 4.75 seconds to 5.92 seconds. Another factor comes from a large portion of small queries that have identical query plans and low execution times, where our approach suffers from the overhead of cardinality estimation. These results show that OmniSketch join estimation can be useful with large performance and intermediate result improvements for star-schema workloads, even if the intermediate result decreases do not fully translate to execution time decreases. For other schema shapes, the applicability of OmniSketch may be limited due to frequent loss of witnesses.

\definecolor{dollarbill}{RGB}{0, 176, 0}
\definecolor{regression}{RGB}{215, 46, 55}
\begin{table}
\caption{End-to-end execution time and intermediate result improvements and regressions for SSB-skew and JOB-light of our approach (Omni) compared to DuckDB.}
\centering
\begin{tabular}{l | r r | r r} 
 \hline
  & \multicolumn{2}{c|}{\textbf{SSB-skew}} & \multicolumn{2}{c}{\textbf{JOB-light}} \\
 \hline
  & \textbf{Omni} & \textbf{DuckDB} & \textbf{Omni} & \textbf{DuckDB} \\
 \hline
 $\sum$ \textbf{Execution time} [s] & 4.61 & 7.57 & 12.46 & 10.65 \\ 
 Max. improvement & \multicolumn{2}{c|}{\textbf{\color{dollarbill}3.19x}} & \multicolumn{2}{c}{\textbf{\color{dollarbill}2.14x}} \\
 Max. regression & \multicolumn{2}{c|}{\textbf{\color{regression}0.98x}} & \multicolumn{2}{c}{\textbf{\color{regression}0.25x}} \\
 \hline
 $\sum$ \textbf{Intermediates} & 367\,M & 2,620\,M & 14,932\,M & 14,467\,M  \\
 Max. improvement & \multicolumn{2}{c|}{\textbf{\color{dollarbill}1077x}} & \multicolumn{2}{c}{\textbf{\color{dollarbill}3.97x}} \\
 Max. regression & \multicolumn{2}{c|}{\textbf{\color{regression}0.86x}} & \multicolumn{2}{c}{\textbf{\color{regression}0.92x}} \\
 \hline
\end{tabular}
\vspace{-0.2cm}
\label{tbl:plan-quality}
\end{table}

\section{Conclusion}
We introduced a new concept for join cardinality estimation with OmniSketches. OmniSketches can be used to estimate the cardinality of multi-join, multi-predicate queries and are especially useful if the data is organized in a star-schema so that we can build secondary sketches on dimension tables. For such a case, our experiments with the SSB-skew benchmark show intermediate result decreases of up to 1,077x and performance improvements of up to 3.19x. In other cases, OmniSketch join estimation may suffer from long cardinality estimation latencies and running out of witnesses. Interesting directions of future work include the applicability for galaxy schemas, cardinality estimation for additional operators, and deriving error bounds for multi-join OmniSketch estimates.

\section*{Acknowledgment}
We gratefully acknowledge funding from the German Federal Ministry of Research, Technology and Space under the grant BIFOLD25B.

\bibliographystyle{IEEEtran}
\bibliography{references}

\end{document}